\documentclass[aps,prc,twocolumn,superscriptaddress,showpacs,amsmath,floatfix,
nofootinbib]{revtex4}
\usepackage{graphicx}
\usepackage{bm}

\begin{document}

\title{
Rotation alignment in neutron-rich Cr isotopes: A probe of deformed
single-particle levels across $N=40$ }

\author{
Yingchun Yang $^{1}$, Yang Sun $^{1,2,3}$, Kazunari
Kaneko $^{4}$, Munetake Hasegawa $^{1,2}$}

\affiliation{ $^{1}$Department of Physics, Shanghai Jiao Tong
University, Shanghai 200240, People's Republic of China \\
$^{2}$Institute of Modern Physics, Chinese Academy of Sciences,
Lanzhou 730000, People's Republic of China \\
$^{3}$Department of Physics and Astronomy, University of Tennessee,
Knoxville, Tennessee 37996, USA \\
$^{4}$Department of Physics, Kyushu Sangyo University, Fukuoka
813-8503, Japan }

\begin{abstract}

Recent experiments have reached the neutron-rich Cr isotope with
$N=40$ and confirmed enhanced collectivity near this sub-shell.  The
current data focus on low-spin spectroscopy only, with little
information on the states where high-$j$ particles align their spins
with the system rotation.  By applying the Projected Shell Model, we
show that rotation alignment occurs in neutron-rich even-even Cr
nuclei as early as spin 8$\hbar$ and, due to shell filling, the
aligning particles differ in different isotopes.  It is suggested
that observation of irregularities in moments of inertia is a direct
probe of the deformed single-particle scheme in this exotic mass
region.

\end{abstract}


\date{\today}
\maketitle

Current nuclear structure studies are devoted to the discussion of
enhanced collectivity in the neutron-rich $pf$-shell nuclei with
neutron-number $N\approx 40$.  One has found strong evidence for
compressed first $2^+$ energy levels and large E2 transitions
linking these and the ground states for several isotopic chains
around the proton magic number $Z=28$, for example, in the Cr
($Z=24$) \cite{Cr64,Aoi09}, Fe ($Z=26$)
\cite{Fe62-64-66Data,Fe66,Fe62-64BE2}, and Zn ($Z=30$)
\cite{Perru06} isotopic chains.  These experimental results support
the early suggestions that near $N=40$, pronounced collectivity
develops corresponding to the formation of a region of deformation
\cite{PRL_Grzywacz98,Fe-deformation,Cr-deformation}.

In the study of neutron-rich nuclei, an important issue is to
understand emerging sub-shell gaps which cause substantial
modifications of the intrinsic shell structure in nuclei with a
neutron excess \cite{Janssens05}.  While information on collective
excitations in low-spin states is useful, a comprehensive knowledge
for these exotic nuclei requires the study of higher-spin states in
which, due to rotation alignment, quasiparticle configurations are
dominant.  For an yrast band consisting of the lowest states for
each angular momentum, the aligning particles carry valuable
information on the deformed single-particle states.  Therefore,
investigations of high-spin spectra can yield knowledge on the
intrinsic shell structure of single-particle levels.

Microscopic calculations have shown that beginning from $N\approx
30$, energy minima with sizable prolate deformations show up for the
neutron-rich Cr isotopes \cite{Egido07}.  In these deformed Cr
isotopes, protons occupy up to the $\pi f_{7/2}$ orbit whilst
neutrons of the $N>28$ isotopes fill in the rest of the $pf$-shell.
With the splitting of single-particle orbits due to deformation, the
proton Fermi level lies between the $f_{7/2}$ orbitals
$\pi[321]3/2^-$ and $\pi[312]5/2^-$, and is also not far from
$\pi[300]1/2^-$.  On the other hand, the down-sloping levels of the
neutron intruder $g_{9/2}$ orbit, $\nu[440]1/2^+$, $\nu[431]3/2^+$,
and $\nu[422]5/2^+$, are found near the neutron Fermi levels, and
therefore, neutrons can easily occupy these orbitals.  Thus when
nuclei rotate, these high-$j$ particles (here, $f_{7/2}$ for protons
and $g_{9/2}$ for neutrons) are among the first to align their
rotation along with the rotation-axis of the system, resulting in
observable effects in the moment of inertia, which correspond to the
phenomenon known as rotation alignment \cite{SS72}.  Thus with
increasing neutron number from $N=30$ towards $40$ and beyond, these
high-$j$ orbits dominate the high-spin behavior of these nuclei.
This qualitative picture is valid also for nuclei with a soft ground
state. Angular-momentum-projected energy-surface calculations show
\cite{PSM_Fe} that as soon as the nuclei begin to rotate,
well-defined shapes in favor of prolate deformation develop.

With the experimental advances, detailed spectroscopic measurements
for neutron-rich nuclei now become possible.  In a very recent work,
Gade {\it et al.} \cite{Cr64} reported their successful experiment
for the neutron-rich isotope $^{64}$Cr by $^9$Be-induced inelastic
scattering, obtaining the first spectroscopy at the $N=40$ subshell
for Cr isotopes.  Data for some lighter Cr isotopes are presently
available \cite{Devlin99,Zhu06,Aoi09}.  In the near future,
fragmentation of a $^{76}$Ge beam may push the experiment to more
neutron-rich regions \cite{Nrich1}.  On the theoretical side,
large-scale shell-model calculations \cite{Caurier02,Honma05,Cr}
have been successful in describing the low-spin spectroscopy of
neutron-rich nuclei.  For example, the spherical shell-model
calculation for Cr isotopes \cite{Cr} including the $g_{9/2}$ orbit
in the model space predicted the first excited $4^+$ energy of
$^{62}$Cr, which was later confirmed by experiment \cite{Aoi09}.
There have been encouraging applications by beyond-mean-field
approaches \cite{Egido07,Gaudefroy09} which can easily handle a
large model space.  Nevertheless, models that either do not allow
sufficient amount of valence particles in the spherical shell model
space or do not build excited quasiparticle configurations in the
deformed models may not be appropriate for discussions of high-spin
physics.

\begin{figure*}[t]
\includegraphics[totalheight=8cm]{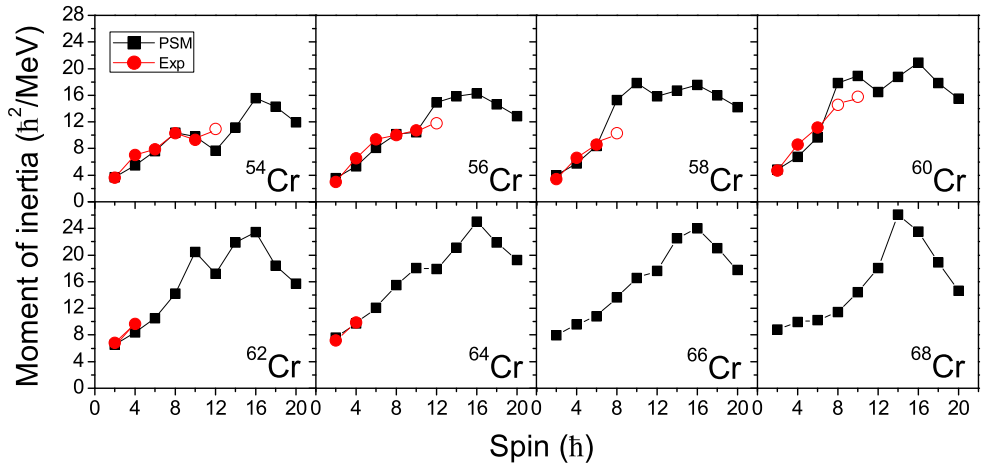}
  \caption{(Color online) Comparison of the calculated moments of inertia
  (filled squares) for the yrast bands in even-even
  $^{54-68}$Cr with the known
  experimental data (filled circles) taken from Refs.
  \cite{Devlin99} ($^{54}$Cr), \cite{Zhu06} ($^{56-60}$Cr),
  \cite{Aoi09} ($^{62}$Cr), and \cite{Cr64} ($^{64}$Cr).
  Note that open circles denote those tentative data reported in these
  publications.}
  \label{fig1}
\end{figure*}

To discuss high-spin states and to further study the deformed
single-particle structure in neutron-rich nuclei, we performed
Projected Shell Model (PSM) \cite{PSM} calculations for
neutron-rich, even-even Cr isotopes with neutron number from 30 to
44, aiming at making predictions ahead of experiment.  The model has
recently been applied to the neutron-rich Fe isotopes \cite{PSM_Fe},
where large $2^+$ state B(E2)'s were predicted for $^{62,64}$Fe and
confirmed later by the measurement \cite{Fe62-64BE2}.  It has also
been employed to study the yrast structure of the Ge nuclei
\cite{PSM_Ge}.  The PSM calculation uses deformed Nilsson
single-particle states \cite{BR85} to build the model basis.  For
the present calculations, the quadrupole deformation parameters for
building the deformed bases are listed in Table \ref{tab:1}.  These
parameters are consistent with the known experimental trend of
increasing deformation towards $N=40$ \cite{Zhu06}, and afterwards a
slightly decreasing collectivity as predicted by spherical shell
model calculations \cite{Cr}.  Pairing correlations are incorporated
into the Nilsson states by a BCS calculation.  The consequence of
the Nilsson-BCS calculations defines a set of quasiparticle (qp)
states corresponding to the qp vacuum $\left|0\right>$.  The PSM
wavefunction is a superposition of (angular-momentum) projected
multi-qp states that span the shell model space
\begin{equation}
\left|\Psi^{\sigma}_{IM}\right> = \sum _{K \kappa}
f^{\sigma}_{IK_\kappa}\,\hat P^I_{MK}\left|\Phi_\kappa \right> ,
\label{wf}
\end{equation}
where $\left|\Phi_\kappa\right\rangle$ denotes the qp-basis,
$\kappa$ labels the basis states and $f^{\sigma}_{IK_\kappa}$ are
determined by the configuration mixing implemented by
diagonalization. $\hat{P}^{I}_{MK}$ is the angular momentum
projection operator \cite{PSM} which projects an intrinsic
configuration onto states with good angular momentum.  As the
valence space for this mass region, particles in three major shells
($N=2,3,4$ for both neutrons and protons) are activated.  The
multi-qp configurations consisting of 0-, 2-, and 4-qp states for
even-even nuclei are as follows:
\begin{equation}
\begin{array}{rl}
\{ \left|0 \right\rangle, a^\dagger_{\nu_i} a^\dagger_{\nu_j}
\left|0 \right\rangle, a^\dagger_{\pi_i} a^\dagger_{\pi_j} \left|0
\right\rangle, a^\dagger_{\nu_i} a^\dagger_{\nu_j} a^\dagger_{\pi_k}
a^\dagger_{\pi_l} \left|0 \right\rangle \} ,
\label{qpset}
\end{array}
\end{equation}
where $a^\dagger_{\nu}$ and $a^\dagger_{\pi}$ are the neutron- and
proton-qp creation operators, respectively, with the subscripts $i$,
$j$, $k$, and $l$ denoting the Nilsson quantum numbers which run
over the orbitals close to the Fermi levels.

\begin{table}
\caption{Input deformation parameters ($\epsilon_2$) used in the
calculation.} \label{tab:1}
\begin{tabular}{c|cccccccc}
\hline\noalign{\smallskip}
Cr & 54 & 56 & 58 & 60 & 62 & 64 & 66 & 68 \\
\noalign{\smallskip}\hline\noalign{\smallskip}
$\varepsilon_2$ & 0.210 & 0.210 & 0.220 & 0.240 & 0.250 & 0.235 & 0.235 & 0.230 \\
\noalign{\smallskip}\hline
\end{tabular}
\end{table}

The PSM calculation employs a quadrupole plus pairing Hamiltonian,
with inclusion of quadrupole-pairing term
\begin{equation}
\hat H = \hat H_0 - {1 \over 2} \chi \sum_\mu \hat Q^\dagger_\mu
\hat Q^{}_\mu - G_M \hat P^\dagger \hat P - G_Q \sum_\mu \hat
P^\dagger_\mu\hat P^{}_\mu . \label{hamham}
\end{equation}
In Eq. (\ref{hamham}), $\hat H_0$ is the spherical single-particle
Hamiltonian which contains a proper spin-orbit force.  The monopole
pairing strengths are taken to be $G_M = \left[G_1 \mp G_2(N -
Z)/A\right] /A$, where "$+$" ("$-$") is for protons (neutrons) with
$G_1=18.72$ and $G_2=10.74$. The quadrupole pairing strength $G_Q$
is assumed to be proportional to $G_M$, the proportionality constant
being fixed to 0.30.

\begin{figure*}
   \centering
  \begin{minipage}[c]{0.68\textwidth}
    \centering
    \includegraphics[width=\textwidth]{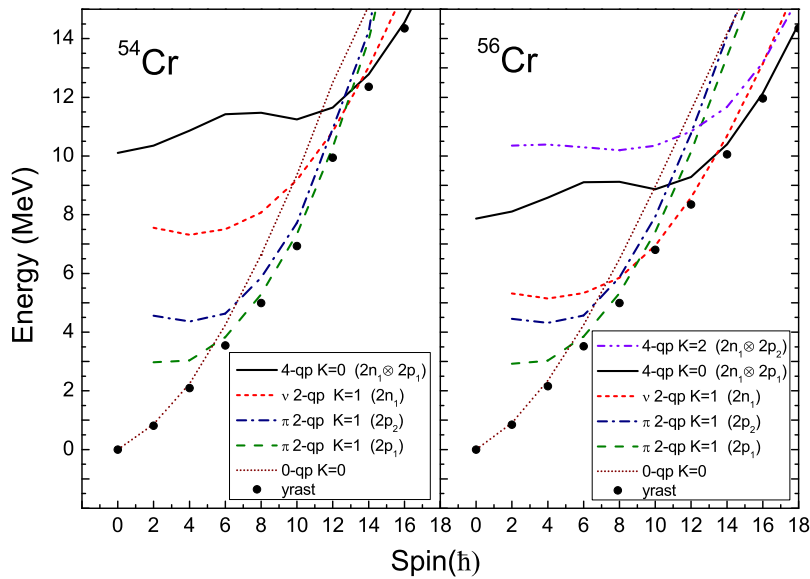}
  \end{minipage}%
  \begin{minipage}[c]{0.3\textwidth}
    \centering
\caption{(Color online) Theoretical band diagrams for $^{54,56}$Cr.
Bands for some important configurations are shown and explained in
Table II.  Note that, to illustrate them clearly, only even-spin
states are plotted to avoid a strong zigzag in curves between even
and odd-spin states. }
    \label{fig2}
  \end{minipage}
\end{figure*}

The moment of inertia (MoI) is a characteristic quantity for the
description of rotational behavior.  In Fig.\ref{fig1}, the
calculated results (filled squares) for the yrast bands in
$^{54-68}$Cr are presented in terms of the MoI (defined as ${\cal
J}(I)=(2I-1)/[E(I)-E(I-2)]$), and compared with the available data
(filled circles).  It is observed that for these isotopes, $\cal J$
increases nearly linearly with spin $I$ for the lowest spin states.
However, irregularities in MoI are seen as early as $I=8$, with the
actual pattern differing in different isotopes.  Interestingly, all
the current data, with the last one or two data points assigned as
tentative in several isotopes, stop at the spins where
irregularities are predicted to occur.  As we shall discuss below,
these irregularities reflect changes in the yrast structure caused
by rotation alignment of nucleons from specific orbitals. Thus by
studying the changes in MoI one can gain valuable information on
deformed single-particle states for this exotic mass region.

By examining the MoI patterns in Fig. 1, one finds that with
increasing spin, $\cal J$ of the two lightest isotopes $^{54,56}$Cr
either bends down or stops rising at spins $I=10$ and 12, and then
shows a rapid rise.  A peak is predicted to occur at $I=16$ for both
isotopes.  Irregularities in $\cal J$ are especially notable in
$^{58,60,62}$Cr, in which the rotational patterns are interrupted
several times.  The first one is seen soon after $I=6$ where a jump
in $\cal J$ is predicted. Later at $I=10$ and 16, two peaks are
predicted to appear.  For the $N\ge 40$ isotopes $^{64,66,68}$Cr,
our calculation suggests an gradual increase in $\cal J$ (with some
small perturbations) up to high spins, until a peak is formed at
$I=14$ or 16.

\begin{table}
\caption{2-qp configurations for neutrons and protons.}
\label{tab:2}
\begin{tabular}{c|c|c|c}
\hline\noalign{\smallskip}
neutron  &   configurations &  protron &  configurations  \\
\noalign{\smallskip}\hline\noalign{\smallskip}
 2n$_{1}$ & $\nu1/2[440] \bigotimes \nu 3/2[431]$ & 2p$_{1}$ & $\pi 3/2[321] \bigotimes \pi 5/2[312]$  \\
\noalign{\smallskip}\hline\noalign{\smallskip}
 2n$_{2}$ & $\nu 3/2[431] \bigotimes \nu 5/2[422]$ & 2p$_{2}$ & $\pi 1/2[300] \bigotimes \pi 3/2[321]$  \\
\noalign{\smallskip}\hline\noalign{\smallskip}
 2n$_{3}$ & $\nu 5/2[422] \bigotimes \nu 7/2[413]$  \\

\noalign{\smallskip}\hline
\end{tabular}
\end{table}

To understand what causes the variations in MoI, we study
theoretical band diagrams for these isotopes.  In the PSM, the
energy of a calculated band $\kappa$ is defined by
\begin{equation}
E_\kappa(I) = {{\left< \Phi_\kappa \right|\hat H \hat P^I_{KK}
\left|\Phi_\kappa\right>}\over{\left< \Phi_\kappa \right|\hat
P^I_{KK} \left|\Phi_\kappa\right>}}, \label{banddiag}
\end{equation}
which is the projected energy of a multi-quasiparticle configuration
in (\ref{qpset}) as a function of spin $I$.  An ensemble of
projected configurations plotted in one figure is called band
diagram \cite{PSM}, in which the rotational behavior of each
configuration as well as its relative energy compared to other
configurations are easily visualized.  Because our deformed basis
states retain axial symmetry, we use $K$ (the projection of angular
momentum on the symmetry axis of the deformed body) to classify the
configurations.  For even-even nuclei, the 0-qp ground band has $K =
0$, whereas a multi-qp band has a $K$ given by the sum of the
Nilsson $K$ quantum numbers of its constituent qp's.  A
superposition of them imposed by configuration mixing gives the
final results, with the lowest one at each spin being the yrast
state.  The study of band crossings in the yrast region can give us
useful messages from which one identifies the most important
configurations for the yrast states.

\begin{figure*}[t]
\centering{}
\includegraphics[totalheight=9.5cm]{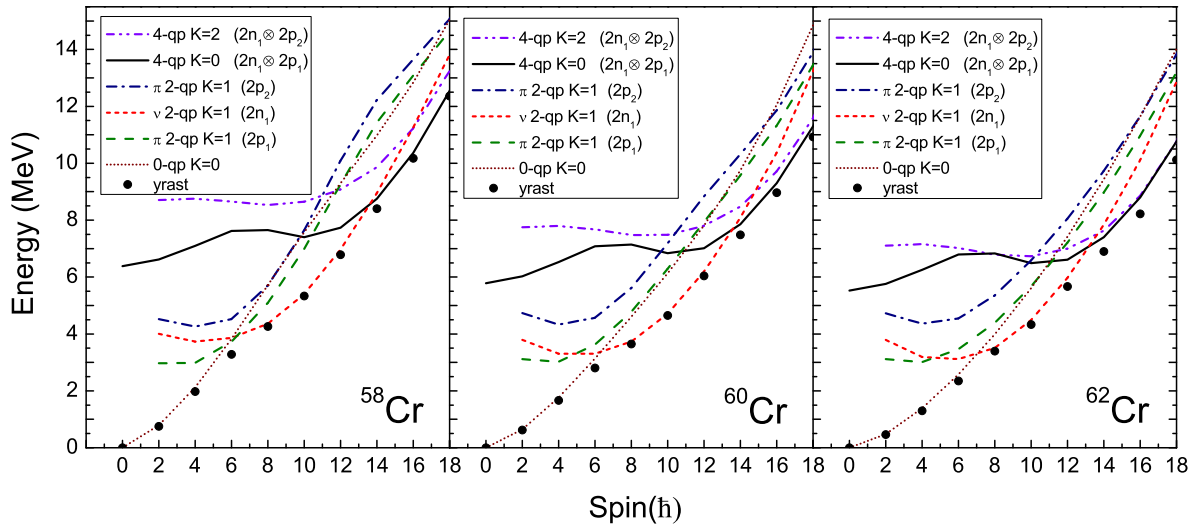}
  \caption{(Color online) Same as Fig \ref{fig2}, but for $^{58,60,62}$Cr. }
  \label{fig3}
\end{figure*}

\begin{figure*}[t]
\includegraphics[totalheight=9.5cm]{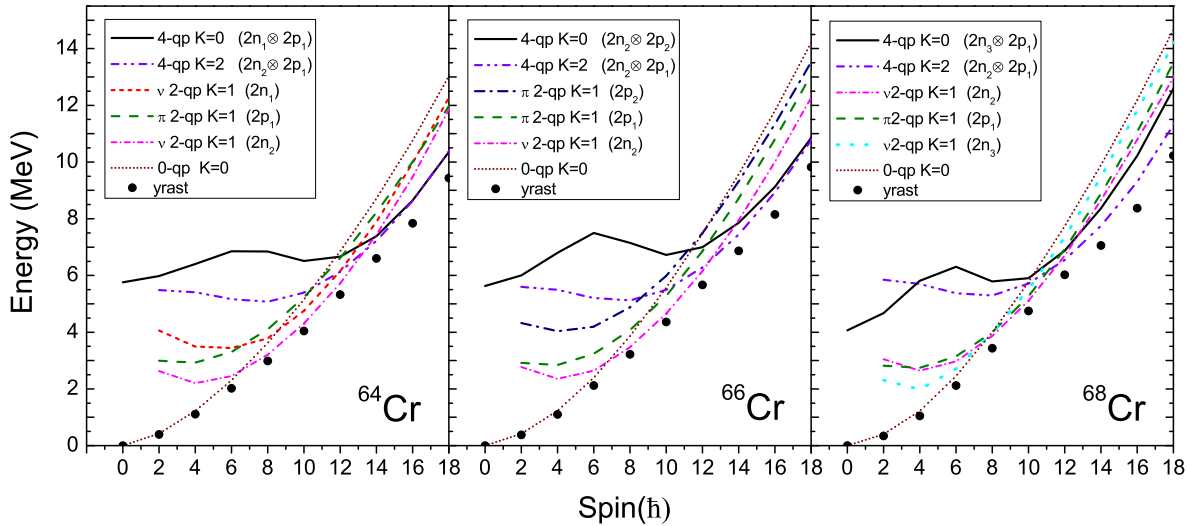}
\caption{(Color online) Same as Fig
\ref{fig2}, but for $^{64,66,68}$Cr. }
  \label{fig4}
\end{figure*}

According to the nature of band-crossings, we may divide the
isotopes into three groups for discussion.  In Figs.
\ref{fig2}-\ref{fig4}, we plot the important configurations that are
used to discuss the MoI variations in Fig. \ref{fig1}.  It can be
seen from the figures that as a nucleus starts rotating, the 0-qp
ground band energy increases and the band quickly enters into the
high energy region.  The 2- and 4-qp bands rise slowly at low spins,
and therefore, can cross the 0-qp band.  To facilitate the
discussion, we term the relevant 2-qp configurations to be 2n$_{1}$,
2n$_{2}$, 2n$_{3}$, 2p$_{1}$, and 2p$_{2}$, with the details listed
in Table \ref{tab:2}.  In the band diagrams, we use different line
styles to distinguish different qp bands, so that one can easily
follow them with increasing neutron number.  In addition, the solid
circles marked as ``yrast" in Figs. \ref{fig2}-\ref{fig4} are the
lowest state at each spin obtained after diagonalization, and these
are the theoretical results compared with the data (see Fig.
\ref{fig1}).

In the two lightest isotopes $^{54,56}$Cr, the neutron $g_{9/2}$
orbitals are far above the Fermi level, and therefore, neutron 2-qp
states are high in energy.  Proton 2-qp states are expected to be
the first to cross the 0-qp band.  In Fig. \ref{fig2}, we indeed
find that the proton 2-qp band (marked as 2p$_{1}$) is the lowest
2-qp band which crosses the ground band at $I=6$.  The crossing is
gentle so that it causes little disturbance in MoI.  Another proton
2-qp band (marked as 2p$_{2}$) exhibits a similar character and
approaches the 2p$_{1}$ band near $I=10$.  After this spin, the two
proton 2-qp bands stay nearly parallel and interact with each other.
The interaction causes the first irregularity in the MoI of
$^{54}$Cr at $I\approx 10$, as seen in Fig. \ref{fig1}.  The neutron
2-qp band 2n$_1$ lies high in energy at low spins; it crosses
however the proton 2-qp bands at $I\approx 10$ in $^{56}$Cr.  In the
spin interval $I=12-14$, the 4-qp band consisting of two 2-qp
states, 2n$_{1}$ and 2p$_{1}$, sharply crosses the 2-qp bands.
After $I\approx 14$, the yrast states are predicted to be of a 4-qp
structure.  The 4-qp band crossing changes the content of the yrast
wave functions, leading to the second irregularity in MoI of
$^{54,56}$Cr, as seen in Fig. \ref{fig1}.

Going toward $N=40$, very irregular MoI's are predicted in Fig.
\ref{fig1} for $^{58,60,62}$Cr.  This is because as neutrons begin
to fill the $g_{9/2}$ shell, the low-$K$ orbitals come close to the
Fermi level.  It is known that for nucleons of low-$K$, high-$j$
orbitals it is easy to align the spins with the rotation.  In Fig.
\ref{fig3}, one sees that although the lowest 2-qp band at low spins
$I<6$ is the proton 2p$_{1}$, it climbs rapidly and soon becomes
unimportant.  The other proton 2-qp band beginning at a higher
energy shows a similar character.  In contrast, the neutron 2-qp
band (marked as 2n$_{1}$), which is coupled by two neutrons in the
low-$K$ $g_{9/2}$ orbitals, is more important because it crosses the
ground band at $I=6$, and dominates the yrast structure in the spin
interval $I=6-14$.  The crossing leads to a jump in MoI predicted at
$I=6 $ or 8, as seen in Fig. \ref{fig1}.  The prediction is at
variance with the current tentative data \cite{Zhu06} which seem to
suggest a more regular rise in MoI.  There are two 4-qp bands in
each isotope shown in Fig. \ref{fig3}, among which the $K=0$ 4-qp
band is more notable, as it crosses the neutron 2-qp band 2n$_1$ at
spin $I\approx 14$, and becomes yrast after that spin. The peak in
MoI at $I\approx 16$ in $^{58,60,62}$Cr is attributed to the
crossing of the 4-qp band.

Band diagrams for the heavier isotopes $^{64,66,68}$Cr are plotted
in Fig. \ref{fig4}.  With increasing neutron number, the 2n$_{1}$
configuration lies far from the Fermi level.  Instead, the other two
neutron 2-qp bands 2n$_{2}$ and 2n$_{3}$ play a role.  Since in the
2n$_{2}$ and 2n$_{3}$ configurations there are higher-$K$ states
which are more coupled, the rotational behavior of 2n$_{2}$ and
2n$_{3}$ is different from 2n$_1$.  2-qp bands with high-$K$ states
exhibit similar rotational behavior as the ground band, and
therefore, crossing of them is always gentle.  This makes an
observable difference in MoI: although 2n$_{2}$ and 2n$_{3}$ bands
cross the ground band at $I\approx 6$, no notable variation in MoI
can been seen in Fig. \ref{fig1}.  The yrast states of high spins
($I>14$) are predicted to be strongly mixed by the $K=2$ 4-qp
configuration. Peaks in MoI at high spins are due to the band
crossing with the 4-qp configurations.

Taking a survey for the band diagrams of Figs.
\ref{fig2}-\ref{fig4}, we clearly see the evolution of the 2-qp and
4-qp band structure along the isotopic chain imposed by shell
fillings.  With increasing neutron number, the role of the g$_{9/2}$
orbit changes with occupation of different $K$ states because the
nucleons that first align their spin vary between the various
isotopes of interest.  For the mass region with less dense
single-particle levels, a distinct behavior in observable quantities
such as the moment of inertia can be detected in neighboring
isotopes.

\begin{figure}[t]
\includegraphics[totalheight=8.2cm]{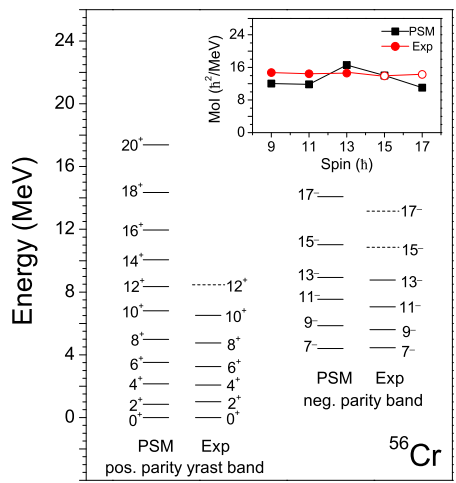}
\caption{(Color online) Comparison of the
calculated energy levels with data \protect\cite{Zhu06} for the
positive-parity yrast band and the negative-parity band in
$^{56}$Cr.  Dashed lines denote those tentatively assigned data. The
inserted plot shows the comparison of the calculated MoI with data
for the negative-parity band.}
  \label{fig5}
\end{figure}

The current experimental data however limit possible comparisons
between calculations and experiment for the high-spin yrast states
and do not allow us to draw strong conclusions about the relevance
of the calculated results.  This is particularly bothersome because
Fig. \ref{fig1} indicates discrepancies between calculations and
experiment at the highest spin states of the experimental bands.
However, there is at least one negative-parity band in $^{56}$Cr
that has been observed \cite{Zhu06} up to high spins.  This band is
of a 2-qp structure with one quasiparticle from the neutron
$g_{9/2}$ orbital. In Fig. \ref{fig5}, we compare two rotational
bands of $^{56}$Cr with data taken from Ref. \cite{Zhu06}.  These
are the lowest positive-parity (yrast) band and the lowest
negative-parity band obtained from the calculation.  The calculation
reproduces nicely the bandhead energy as well as the rotational
feature of the negative-parity band.  The inserted MoI plot in Fig
\ref{fig5} however indicates that the calculation exaggerates the
proton $f_{7/2}$ alignment at $I=13$ with a larger peak in MoI than
what the data show.  This suggests that one must be cautious about
the quantitative details of the prediction in Fig. \ref{fig1}
although we do not expect that small discrepancies can change the
alignment picture discussed in the paper.

As the entire discussion in the present work depends on the
single-particle states, it is important to comment on the deformed
Nilsson scheme employed in the model basis.  The Nilsson parameters
used for the present discussions were fitted a long time ago to the
stable nuclei \cite{BR85}.  It has been shown that the standard
Nilsson parameters may need adjustments when they are applied to
proton- or neutron-rich regions \cite{Zhang98,Sun00}.  For
neutron-rich nuclei with considerable neutron excess, the usage of
the standard parameter set for these nuclei needs to be validated.
It will not be surprising if the Nilsson parameters used here would
require a modification.  Thus future experiments for high-spin yrast
spectra will be compared with the present predictions and serve as
guidance to modify the deformed single-particle scheme.

In conclusion, the present study has highlighted the role of the
proton $f_{7/2}$ and neutron $g_{9/2}$ orbits that are involved in
the discussion of neutron-rich nuclei.  As we have studied in
detail, excitation to these orbits leads to pronounced observation
effects at high spins where variations in the moment of inertia are
explained in terms of rotation alignment of the high-$j$ particles.
As the discussion is based crucially on the deformed single-particle
scheme, we emphasize that any future experimental confirmation or
refutation of our predictions will be valuable information, which
can help to pin down the single-particle structure in this
neutron-rich mass region.

Research at SJTU was supported
by the Shanghai Pu-Jiang scholarship, the National Natural Science
Foundation of China under contract No. 10875077, the Doctoral
Program of High Education Science Foundation under grant No.
20090073110061, and the Chinese Major State Basic Research
Development Program through under grant No. 2007CB815005.


\baselineskip = 14pt
\bibliographystyle{unsrt}


\end{document}